\documentclass[aps,prl,preprint,groupedaddress]{revtex4-1}

\usepackage[]{graphicx} 
\usepackage{amsmath}	
\usepackage{braket}
\usepackage{bm}
\newcommand{\nucl}[2]{{}^{#1}\mathrm{#2}}
\newcommand{\nn}{\nonumber}
\newcommand{\CO}{\nucl{12}{C} + \nucl{16}{O}}

\begin{document}


\title{
$^{\bf 12}{\bf C}\bf +{}^{\bf 16}{\bf O}$ molecular resonances at  deep sub-barrier energy}

\author{Yasutaka Taniguchi}
\email{taniguchi-y@di.kagawa-nct.ac.jp}
\affiliation{Department of Information Engineering, National Institute of Technology (KOSEN), Kagawa College, Mitoyo, Kagawa 769-1192, Japan}
\author{Masaaki Kimura}
 \email{masaaki@nucl.sci.hokudai.ac.jp}
 \affiliation{Department of Physics, Hokkaido University, Sapporo, Hokkaido 060-0810, Japan}
 \affiliation{Nuclear Reaction Data Centre, Hokkaido University, Sapporo, Hokkaido 060-0810, Japan}

\date{\today}

\begin{abstract}
The existence of  $\nucl{12}{C} + \nucl{16}{O}$ molecular resonances at sub-barrier energy has
been a significant problem in nuclear astrophysics because they strongly affect the 
$\nucl{12}{C} + \nucl{16}{O}$ fusion reaction rate in  type Ia supernovae and heavy stars. 
However, experimental surveys have been limited to 4~MeV and cannot access the
deep sub-barrier energy due to a very small fusion cross section. 
Here we predict a couple of resonances with $J^\pi=0^+$, $2^+$, and $4^+$ in the deep sub-barrier
energy based on the antisymmetrized molecular dynamics calculation that reproduces the known
resonances and low-lying spectrum of $^{28}{\rm Si}$. 

\end{abstract}

\maketitle
{\it Introduction.---}
Understanding  the origin of matter is a significant challenge in nuclear and
astrophysics fields. The $\nucl{12}{C} + \nucl{16}{O}$ fusion reaction at thermal energy is critical for nucleosynthesis in type Ia supernovae and heavy stars.  The ignition of type Ia supernovae is
carbon and oxygen burning, and this strongly influences the abundance of elements in the $14
\lesssim Z \lesssim 20$ region \cite{Mart_nez_Rodr_guez_2017}.  If they exist, 
$\nucl{12}{C} + \nucl{16}{O}$ molecular resonances close to the Gamow window increase the cross-section in order of magnitude.
 Therefore, searching for molecular
resonances and determining  their resonance parameters are of primary importance for understanding of the reaction rate.

Experimentally, resonances at 3.9--4.0 and 4.4~MeV have been observed by low-energy $\CO$ fusion reaction, and an increased $\CO$ fusion
reaction rate has been reported previously \cite{PhysRevC.96.045804}.
In addition, resonances at a higher energy region have also been observed by radiative capture reactions
\cite{PhysRevC.63.034315,PhysRevC.89.014305} and other reactions \cite{Cindro1981}. Thus, many
experiments have been performed to identify the resonances at thermal energy; however, resonances
at deep sub-barrier energy are still unclear. 

For this problem, theoretical studies are far from satisfactory although a number of cluster model
calculations have been performed. For example, Baye {\it et al.} performed a resonating group method (RGM)
calculation for $\CO$ system and obtained two 
positive parity rotational bands \cite{Baye1976445,Baye1977176}, one of which is just above the  $\CO$
threshold and may have some influence on the reaction rate. Kat\=o {\it et al.} performed an
orthogonality condition model (OCM) calculation including both elastic and inelastic channels
($^{12}{\rm C}(0^+_1, 2^+_1, 4^+_1)$ states) \cite{doi:10.1143/PTP.74.1053}. 
They demonstrated the existence of a much greater number of the molecular states compared to the
RGM calculation, likely owing to elastic and inelastic channel mixing. Kanada-En'yo {\it et al.}
reported the  existence of three positive-parity $\CO$  molecular bands based on the
antisymmetrized molecular dynamics (AMD) calculation without a priori assumption of the $\CO$
cluster structure \cite{KANADAENYO20043}. Similar to the RGM calculation, one of molecular bands is close to the $\CO$
threshold. Thus, many calculations have been performed; however, the number of molecular
resonances and their energies are quite different according to theoretical models. This
uncertainty comes from the ambiguity of effective nucleon-nucleon interactions and limitation of
their model space. 

To overcome this problem, we performed the AMD calculation using the effective Gogny D1S
 interaction \cite{Berger1991365}. 
 This calculation has already been performed for the low-lying states of $^{28}{\rm Si}$ 
 (the composite system of $\CO$ cluster) \cite{PhysRevC.80.044316}. It reasonably described both
 the oblately deformed ground-state band and a prolate normal-deformed (ND) band. 
 Furthermore, it predicted a superdeformed (SD) band having large overlap with the $\alpha+^{24}{\rm Mg}$
cluster configuration, which was subsequently identified experimentally \cite{PhysRevC.86.064308}. In
this study, we extend this calculation to the highly excited states to identify  $\CO$ molecular
 resonances. To cover the broad model space required to study $\CO$ resonances, we 
introduce a constraint on the inter-cluster distances which enables us to describe various
cluster configurations.


{\it Theoretical framework.---}
We employ the AMD framework \cite{PTP.93.115,PhysRevC.56.1844,PhysRevC.69.044319,Kimura2016} with the Gogny
D1S effective interaction \cite{Berger1991365}, which successfully described the low-lying and
superdeformed states of $^{28}{\rm Si}$ \cite{PhysRevC.80.044316}.  The basis wave function of AMD is
a parity-projected  Slater determinant of the single-particle wave packets,  
\begin{align}
 \Phi^\pi_\mathrm{int} =& \frac{1+\pi P_x}{2}\mathcal{A}\set{\varphi_1,...,\varphi_A},\\
 \varphi_i=&\prod_{\sigma=x,y,z}\left(\frac{2\nu_\sigma}{\pi}\right)^{1/4}\exp
 \set{-\nu_\sigma\left(r_\sigma - Z_{i\sigma}\right)^2}\nn\\
 &\times  \left(\alpha_i\ket{\xi_\uparrow}+\beta_i\ket{\xi_\downarrow}\right)
 \times \left(\ket{p}\ \text{or}\ \ket{n}\right),
\end{align}
where each wave packet $\varphi_i$ has the following parameters: Gaussian centroid $\mathbf{Z}_i$,
 width parameter $\bm \nu$ and spin direction $\alpha_i$ and $\beta_i$. 
Note that isospin part is fixed to either of proton ($p$) or neutron ($n$).
Here, $\bm \nu$ is a real vector, and the other parameters are complex numbers. 
All parameters are determined by the energy variation with two different
constraints. The first is a constraint on the quadrupole deformation and the second is on the
inter-cluster distance $d$ between $\nucl{12}{C}$-${}^{16}{\rm O}$ clusters, and between
$\alpha$-${}^{24}{\rm Mg}$ clusters. The constraint on the inter-cluster distance naturally yields
the clustered basis wave functions handling the cluster polarization effect
\cite{PTP.112.475}.  After energy variation, the basis wave functions are projected to the
eigenstates of the angular momentum and superposed to diagonalize the nuclear Hamiltonian. 

The obtained wave functions of ${}^{28}{\rm Si}$ are an admixture of the 
${}^{12}{\rm C}$+${}^{16}{\rm O}$ cluster,  $\alpha$+${}^{24}{\rm Mg}$ cluster and non-cluster
wave functions. To identify the  ${}^{12}{\rm C}$+${}^{16}{\rm O}$ molecular states, we calculated 
the reduced width amplitude (RWA), which is the probability amplitude to find the clusters at
the inter-cluster distance $a$: thus, the RWA is a good measure for clustering. It is defined
as follows: 
\begin{align}
y_l(a) = \sqrt{\frac{28!}{12! 16!}}
 \Braket{\frac{\delta(r-a)}{r^2}\Phi_{^{12}{\rm C}}\Phi_{^{16}\mathrm{O}}Y_{l}(\hat r)|
  \Psi_{^{28}{\rm Si}}^{l}},\label{eq:rwa1}
\end{align}
where the bra state is the reference cluster state, in which the $^{12}{\rm C}$ 
($\Phi_{\nucl{12}{C}}$) and $^{16}{\rm O}$ ($\Phi_{\nucl{16}{O}}$) 
clusters are coupled to angular momentum $l$ with inter-cluster distance $a$.
 The ket state is the wave function of $^{28}{\rm Si}$ calculated by AMD. 
Here, $\Phi_{\nucl{12}{C}}$ and $\Phi_{\nucl{16}{O}}$ are the $(0s)^4(0p_{3/2})^8$
\cite{PhysRevC.87.054334} and $(0s)^4(0p)^{12}$ configurations of the harmonic oscillator
potential, respectively.   
In practical calculation, Eq.~(\ref{eq:rwa1}) was estimated using an approximate
method \cite{Kanada-En'yo01072014}.

{\it Results and discussions.---} 
\begin{figure}[tbp]
  \includegraphics[width=0.5\textwidth]{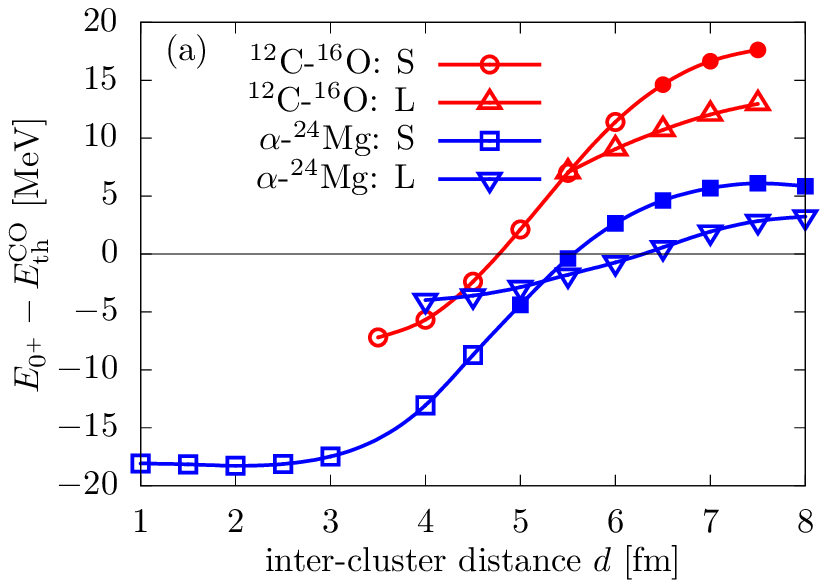}
  \includegraphics[width=0.5\textwidth]{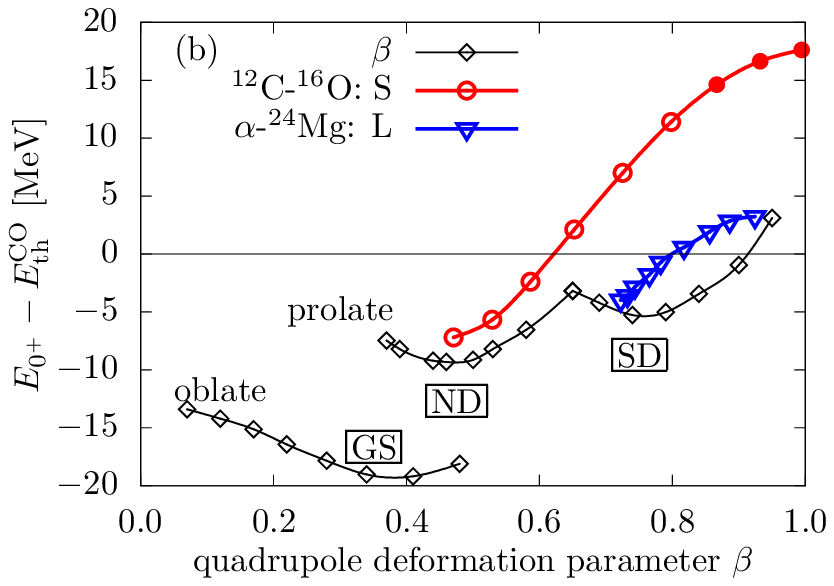}
 \caption{ Energy curves of the $J^\pi=0^+$ state obtained by energy variation with constraints.
 (a) Energy curves for the ${}^{12}{\rm C}+{}^{16}{\rm O}$ and $\alpha + {}^{24}{\rm Mg}$ 
 cluster systems as functions of inter-cluster distance. Two different configurations
 of clusters are calculated (denoted S- and L-type).
 (b) Energy curve obtained by the constraint on the quadrupole deformation as a function of 
 deformation parameter $\beta$ compared to the energy curves for the cluster
 configurations shown in panel (a). } 
 \label{energy}
\end{figure}
Figure~\ref{energy} (a) shows the energy curves obtained by the energy variation with the
constraint on inter-cluster distance $d$. Energies are measured from $\CO$ threshold energy that
is determined by the energy variation of the AMD wave functions of $\nucl{12}{C}$ and
$\nucl{16}{O}$ clusters after parity projection.  For each of the $\nucl{12}C + \nucl{16}{O}$ and $\alpha + \nucl{24}{Mg}$ cluster 
systems, two different configurations (denoted S-type and L-type) were obtained.
In the S-type
configuration, the spherical clusters ({\it i.e.}, $\nucl{16}{O}$ and $\alpha$) are located on the short axis of
the deformed clusters ({\it i.e.}, $\nucl{12}{C}$ and $\nucl{24}{Mg}$).
In the L-type configuration, they are located on the
long axis.  At larger inter-cluster distance ($d > 5.5$ fm), the L-type configurations have
smaller energy due to the larger overlap between clusters.
The S-type cluster wave functions with long inter-cluster distance were prepared by shifting
cluster position in the S-type wave functions with $d = 6$ and 4.5~fm for the
$\nucl{12}{C}$-$\nucl{16}{O}$ and $\alpha$-$\nucl{24}{Mg}$ wave functions, respectively. 
The figure also shows that that the 
Coulomb barrier height for the $\nucl{12}C + \nucl{16}{O}$ system was estimated as approximately
10~MeV above the ${}^{12}{\rm C}+{}^{16}{\rm O}$ threshold. This estimation reasonably coincides
with a simple formula, $V_c=\alpha Z_1 Z_2/(R_1+R_2)$, where $\alpha$, $Z_1$, $Z_2$, $R_1$ and
$R_2$ denote the fine structure constant, the charges of the clusters, and the radii of the
clusters, respectively.  

Panel (b) shows the energy curve obtained by the constraint on the quadrupole
deformation parameter $\beta$. It has three minima denoted GS, ND, and SD, which are the dominant 
components of the ground-state, ND, and SD bands, respectively \cite{PhysRevC.80.044316}. 
The figure also shows the energy curves for the S-type $\nucl{12}C + \nucl{16}{O}$ 
and L-type $\alpha + \nucl{24}{Mg}$ cluster systems which are same as those shown in panel
(a), but plotted as the functions of $\beta$. The energy of these cluster configurations become
close to the ND and SD minima at shorter inter-cluster distance, which indicates the large overlap
between the ND and $\CO$ cluster configurations, and between the SD and $\alpha+{}^{24}{\rm Mg}$
cluster configurations. In fact, it was found that the wave functions 
at the ND and SD configurations have non-negligible contributions to the molecular states.  

\begin{figure}[tbp]
 \includegraphics[width=0.5\textwidth]{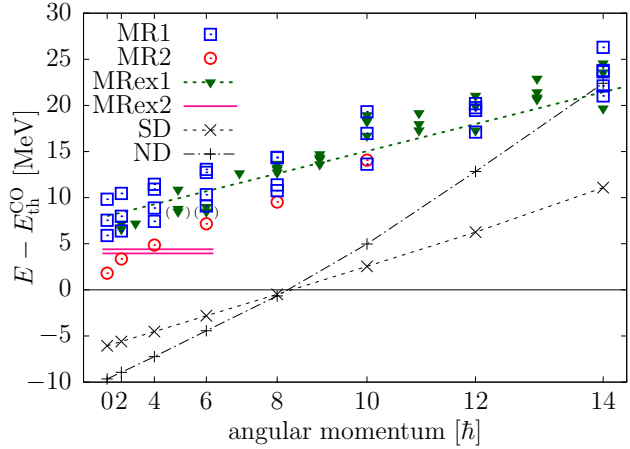}
 \caption{ Level scheme of the excited states with large $\CO$ RWA.
 MR1 and MR2 are the calculated $\CO$ molecular resonances which, respectively, correspond to the
 observed resonances denoted MRex1 \cite{PhysRevC.63.034315,PhysRevC.89.014305,Cindro1981} and
 MRex2 \cite{PhysRevC.96.045804}.
 ND and SD denote normal deformed and superdeformed bands, respectively.
 }
 \label{level}
\end{figure}

\begin{figure}[tbp]
 \includegraphics[width=0.5\textwidth]{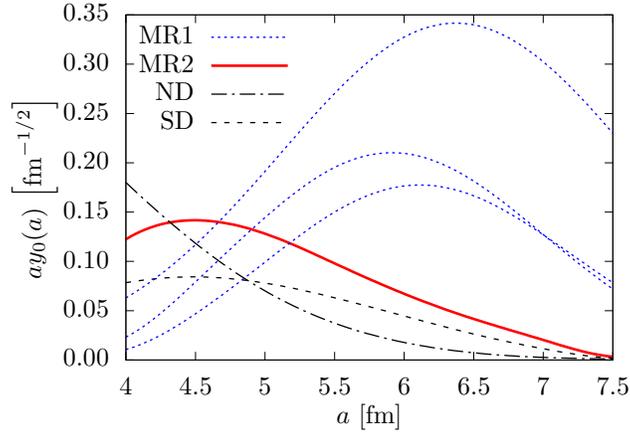}
 \caption{The RWA of the $J^\pi=0^+$ states of MR1, MR2, ND, and SD bands for $\CO$ clustering.
 }
 \label{fig:rwa}
\end{figure}

\begin{figure}[tbp]
 \includegraphics[width=0.5\textwidth]{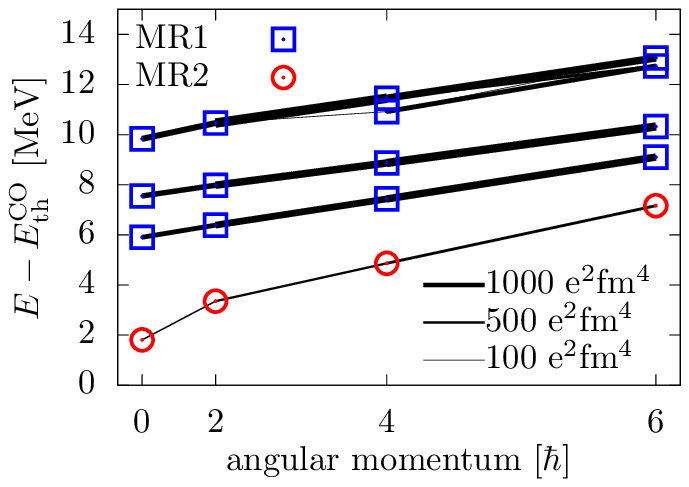}
 \caption{
 $B(E2)$ transition strengths of MR1 and MR2 bands.
 The Widths of lines connecting those states are proportional to $B(E2)$ values.
 }
 \label{fig:be2}
\end{figure}

The GCM calculation was performed by superposing these basis wave functions. As discussed in
Ref.~\onlinecite{PhysRevC.80.044316}, it reasonably described the low-lying states, {\it i.e.},
the oblate-deformed ground band and prolate ND band. It also predicted the SD
band, which was confirmed experimentally \cite{PhysRevC.86.064308}.  
Following these results, in this study, we focus on the ${}^{12}{\rm C}+{}^{16}{\rm O}$
molecular states, which are of astrophysical interest. For this purpose, we selected the
excited states with large RWA for ${}^{12}{\rm C}+{}^{16}{\rm O}$ clustering.
 Figure~\ref{level} summarizes the obtained levels classified as the 
${}^{12}{\rm C}+{}^{16}{\rm O}$ molecular states. They are denoted ND, SD, MR1 and MR2 bands, and 
their properties are as follows: 
\begin{enumerate}
 \item[(1)] MR1: A group of the highly excited states, which lie at 5 to 10~MeV in the case
	    of the $J^\pi=0^+$ states, with huge RWA that have a peak at $a=6$--7~fm
	    (Fig.~\ref{fig:rwa}). We identified these states as molecular resonances (MR1).  
	    The energy and moment-of-inertia of MR1 agree with the molecular resonances
	    observed by the $\nucl{12}{C}(\nucl{16}{O}, \nucl{28}{Si}^\ast)$ and 
	    $\nucl{24}{Mg}(\alpha, \nucl{12}{C})\nucl{16}{O}$ reactions  (triangles and
	    dotted line in Fig.~\ref{level}) fairly well. Note that no other
	    microscopic calculations have explained the observed molecular resonance.

	    The Coulomb barrier height is approximately 10~MeV; thus, these states are
	    considered as the barrier-top resonances. Due to the pronounced clustering, the
	    $B(E2)$ strength between the $J^\pi \leq 6^+$ states are in the order of
	    $10^3~\mathrm{e}^2\rm fm^4$ and much stronger than other states (Fig.~\ref{fig:be2}).  
	    The $B(E2)$ strengths also suggests that the MR1 states with $J^\pi \leq 6^+$ can be
	    classified into three rotational bands. However, we also found that the inter-band
	    transitions become very strong for the $J^\pi > 6^+$ states and the band assignment 
	    is rather ambiguous for high-spin states.

 \item[(2)] MR2: In addition to MR1, we found another molecular band which we denote MR2.
	    MR2 is built on the $0^+$ state just above the threshold (1.8~MeV).  The energy of
	    this band is significantly less than the Coulomb barrier (deep 
	    sub-barrier); thus, the RWA of this band is less than the MR1 states but greater than the ND
	    and SD bands (Fig.~\ref{fig:rwa}). Furthermore, the peak position of RWA is at shorter
	    distance ($a=4.5$~fm), which suggests a compact $\CO$ resonances. Reflecting
	    this,  the  $B(E2)$ strengths and  moment-of-inertia are smaller than MR1.
	    The energy of this band is within the Gamow window of the Carbon burning process;
	    therefore, it is likely 
	    to have an impact on the $^{12}{\rm C}+{}^{16}{\rm O}$ fusion reaction rate in the
	    stellar environment. Interestingly, Fang {\it et al.} reported the existence of the
	    resonances at 3.9--4.0 and 4.4~MeV in a low-energy $\CO$ fusion reaction
	    experiment \cite{PhysRevC.96.045804}, which reasonably coincide with the $2^+$ and $4^+$
	    states of MR2 band. A more detailed study of this band will be important to understand
	    of the $^{12}{\rm C}+{}^{16}{\rm O}$ fusion reaction rate, which affects the abundance
	    of elements in the $14 \lesssim Z \lesssim 20$ region \cite{Mart_nez_Rodr_guez_2017}.  

 \item[(3)] The ND band built on the prolately deformed $0^+_3$ state has overlap with the
	    S-type ${}^{12}{\rm C}+{}^{16}{\rm O}$ configuration; however, the RWA of this band is not
	    as large as the MR1 and MR2 due to the deeper binding. In this band, the cluster
	    configuration is considerably distorted by the formation of the prolately deformed
	    mean-field.   Nevertheless, this band may be classified as the lowest
	    Pauli-allowed ${}^{12}{\rm C}+{}^{16}{\rm O}$ molecular band because no other low-lying states
	    have large ${}^{12}{\rm C}+{}^{16}{\rm O}$ RWA.
 \item[(4)] Unexpectedly, the SD band, which is predominated by the L-type
	    $\alpha+{}^{24}{\rm Mg}$ cluster configuration, also has non-negligible 
	    ${}^{12}{\rm C}+{}^{16}{\rm O}$ RWA that is even greater than the ND band at the long inter-cluster
	    distance as shown in  Fig.~\ref{fig:rwa}, which shows the strong mixing of the $\alpha+{}^{24}{\rm Mg}$ 
	    and ${}^{12}{\rm C}+{}^{16}{\rm O}$ cluster channels.
	    Note that this mixing characteristic reasonably agrees with the observation, {\it i.e.,} it has been reported 
	    that the observed SD states are strongly populated in all of the  
	    $^{12}{\rm C}(^{20}{\rm Ne},\alpha)^{28}{\rm Si}$\cite{Kubono1986461,PhysRevC.86.064308}, 
	    ${}^{24}{\rm Mg}(^{6}{\rm Li},d)^{28}{\rm Si}$\cite{PhysRevC.24.2556} and 
	    ${}^{24}{\rm Mg}(\alpha,\gamma)^{28}{\rm Si}$\cite{Brenneisen1995A,Brenneisen1995B,Brenneisen1995C} reactions. 
	    Thus, the present calculation provides further evidence for the SD band in $^{28}{\rm Si}$
	    and its cluster nature.   
\end{enumerate}
In short, we predict the existence of the deep sub-barrier $^{12}{\rm C}+{}^{16}{\rm O}$ molecular 
states based on the AMD calculation which reasonably reproduces the known low-lying states,
the SD band and  $^{12}{\rm C} + ^{16}{\rm O}$ molecular resonances.


{\it Conclusions.---}
In summary, the $\nucl{12}{C} + \nucl{16}{O}$ molecular resonances were investigated using the AMD
framework. By superposition of the wave functions of the $\nucl{12}{C}$-$\nucl{16}{O}$ 
and $\alpha$-$\nucl{24}{Mg}$ clusters and non-cluster states,  the low-lying and superdeformed
bands and the known $\CO$ molecular resonances were reproduced reasonably. Furthermore, the
calculation predicted the extremely low-energy $\CO$ resonances close to the $\CO$ threshold. 
Some of these $\CO$ resonances are the candidates of the resonances observed by the low-energy
$\CO$ fusion reactions and are expected to have great impact on stellar reactions.

\begin{acknowledgments}
This work was supported by the Hattori Hokokai Foundation Grant-in-Aid for Technological and
Engineering Research, a grant for the RCNP joint research project, the collaborative
research program 2018/2019 at Hokkaido University, and JSPS KAKENHI Grant No. 19K03859.
Numerical calculations were performed using Oakforest-PACS at the CCS, University of Tsukuba, and  XC40 at  YITP, Kyoto University.  
\end{acknowledgments}

\bibliography{MR12C16O_v0}

\end{document}